\documentclass[journal,twocolumn]{IEEEtran}   %ÔËÐÐ1ÏÂ
\usepackage{graphicx}
\usepackage{epstopdf}
\usepackage{amssymb}
\usepackage{amsfonts}
\usepackage{amsmath}
\usepackage{booktabs}
\usepackage{algorithm}
\usepackage{algorithmic}
\usepackage{subeqnarray}
\usepackage{cases}
\usepackage{amsmath,amsfonts}
\usepackage{bm}
\usepackage{subfigure,amsmath,amssymb,cite}
\usepackage{stfloats}
\usepackage{color}
\usepackage{booktabs}

\usepackage{float}
\ifCLASSINFOpdf
\else
\fi
\begin{document}

\title{Enhanced-rate Iterative Beamformers for Active IRS-assisted Wireless Communications}

\author{Yeqing Lin,~Feng Shu,~Rongen Dong,~Riqing Chen,~Siling Feng,\\
~Weiping Shi,~Jing Liu,~and Jiangzhou Wang,\emph{ Fellow, IEEE}

\thanks{This work was supported in part by the National Natural Science Foundation of China (Nos.U22A2002, 62071234 and 61972093), the Hainan Province Science and Technology Special Fund (ZDKJ2021022), Fujian University Industry University Research Joint Innovation Project (No. 2022H6006), and the Scientific Research Fund Project of Hainan University under Grant KYQD(ZR)-21008. (Corresponding authors: Feng Shu, Rongen Dong and Siling Feng)}
\thanks{Feng Shu is with the School of Information and Communication Engineering, Hainan University, Haikou, 570228, China, and also with the School
of Electronic and Optical Engineering, Nanjing University of Science and Technology, Nanjing, 210094, China (e-mail: shufeng0101@163.com)}
\thanks{Yeqing Lin, Rongen Dong, Siling Feng and Jing Liu are with the School of Information and Communication Engineering, Hainan University, Haikou, 570228, China. }
\thanks{Riqing Chen is with the Digital Fujian Institute of Big Data for Agriculture, Fujian Agriculture and Forestry University, Fuzhou 350002, China (e-mail: riqing.chen@fafu.edu.cn)}
\thanks{Weiping Shi is with the School of Electronic and Optical Engineering, Nanjing University of Science and Technology, 210094, China.}
\thanks{Jiangzhou Wang is with the School of Engineering, University of Kent, Canterbury CT2 7NT, U.K. (Email: j.z.wang@kent.ac.uk)}

% <-this % stops a space

}

\maketitle

\begin{abstract}
 Compared to passive intelligent reflecting surface (IRS), active IRS is viewed as  a more efficient promising technique to combat the double-fading impact in IRS-aided wireless network. In this paper, in order to boost the achievable rate of user in such a wireless network,  three enhanced-rate iterative beamforming methods are proposed by designing the amplifying factors  and the corresponding phases at active IRS.  The first method, maximizing the simplified signal-to-noise ratio (Max-SSNR) is designed by omitting the cross-term in the definition of rate. Using the Rayleigh-Ritz (RR) theorem, Max-SSNR-RR is proposed to iteratively optimize the norm of beamforming vector and its associated normalized vector. In addition,  generalized maximum ratio reflection (GMRR) is presented with a closed-form expression, which is motivated by the maximum ratio combining. To further improve rate, maximizing SNR (Max-SNR) is designed by fractional programming (FP), which is called Max-SNR-FP. Simulation results show that  the proposed three methods make an obvious rate enhancement over Max-reflecting signal-to-noise ratio (Max-RSNR), maximum ratio reflection (MRR), selective ratio reflecting (SRR), equal gain reflection (EGR) and passive IRS, and are in increasing order of rate performance as follows: \textcolor{blue}{Max-SSNR-RR, GMRR, and Max-SNR-FP}.
\end{abstract}
\section{Introduction}

Recently, due to its low circuit cost and power consumption,  intelligent reflecting surface (IRS) is becoming an extremely hot research topic and a very promising technique for future wireless networks like B5G and 6G \cite{Wuqingqing2019}.  It has the following major advantages: extend coverage \cite{yang2020coverage}, improve rate \cite{Wuqingqing2020}, enhance security \cite{shiweiping2021secure}, strengthen covert communication \cite{zhouxiaobo2022}, and increase spatial degrees of freedom (DOF) \cite{Lijiayu2021,do2022line}. To achieve the above goals,  IRS can be diversely combined with directional modulation networks \cite{Lijiayu2021}, multiple-input multiple-output \cite{Pancunhua2020}, and relay networks \cite{Wangxuehui2022}. A mobile IRS has been applied to undermine communications to remove the wireless blind zone and improve the wireless communication connectivity in such a situation.

In IRS-aided single-input single-output (SISO) system \cite{yang2020coverage}, the authors derived outage probability and channel ratio gain probability expressions and then used them to analyze the coverage of the IRS,  which showed that IRS provides greater coverage, and improves signal-to-noise ratio gain. \cite{shiweiping2021secure} studied the security of IRS-assisted multi-group multicast communication system, and proposed alternative optimization schemes based on semidefinite relaxation and second-order cone programming. The simulation results  showed that the two proposed schemes can greatly improve the system performance.
%In IRS-aided covert communication \cite{zhouxiaobo2022}, the authors jointly designed the the reflection coefficient of IRS and the transmission power of base station (BS), penalty successive convex approximation algorithm and a low-complexity two-stage algorithm were proposed.
In IRS-assisted beamforming and broadcasting \cite{Tangwankai2020wireless}, the authors derived the free-space path loss, which was related to the IRS distance from the transmitter/receiver IRS, the size of the IRS, the near field/far field and the radiation pattern of the antenna and cell. To analyze the impact of discrete phase shifters on the performance of IRS-aided wireless network system, based on the law of large numbers and Rayleigh distribution, the performance loss analysis was presented under the line-of-sight channels and Rayleigh channels respectively \cite{dong2022performance}.

In addition, IRS can be applied in various scenarios, such as: multi-user multiple-input single-output communication system \cite{guo2020weighted}, and multi-IRS-assisted cell-free network \cite{zhang2021ajoint}, respectively.  In \cite{guo2020weighted}, the authors applied  fractional programming (FP)  and the non-convex block coordinate descent to obtain a smooth solution. In \cite{zhang2021ajoint},  FP  was applied by transforming the non-convex function through a multidimensional complex quadratic transform, thus optimizing the beamforming vectors through an alternating optimization structure.

However, due to the double fading effect of the reflecting channel link between the transmitter and the user, passive IRS helps to achieve a limited gain in rate performance compared to the case without IRS. In order to compensate for the serious dual-path loss in the cascade channel, the passive IRS need to be equipped with thousands or ten thousands  elements to achieve significant passive beamforming gain \cite{najafi2020physics}.  Moreover, for the traditional passive structure, IRS can only adjust the phase of the incident signal, which limits the gain brought by beamforming.  As active IRS emerges, the double-fading effect can be broken completely due to the fact that it is capable of amplifying and reflecting  the incident signal simultaneously. In the active IRS-assisted multi-user system \cite{xu2021resource}, the reflection matrix of the IRS and the beamforming vector of the BS were jointly optimized with the objective of minimizing the transmission power of the BS. To deal with the nonconvex design problem, the authors proposed the bilinear transformation and inner approximation methods.  In \cite{zhi2022active}, the authors derived the optimal power distribution between the transmit signal power of the BS and the output signal power of the IRS, and the results showed that active IRS was superior if the power budget was not very small and the number of IRS components was not very large. In an active IRS-assisted single-cell wireless network \cite{li2022active}, a hybrid Gamma distribution was applied to obtain the average signal-to-noise ratio at user fading. With a fixed budget and a small number of reflected cells, active IRS obtained higher throughput than passive IRS.

In \cite{Dailinglong2021active}, a comparison between passive and active IRSs was made to show that the latter  may harvest a dramatic rate gain over the former in typical scenarios. In the single-user case,  in \cite{Dailinglong2021active}, the authors proposed a equal-gain reflecting (EGR) beamformer with phase alignment with the corresponding product channel and equal amplifying factor. Obviously, this method has not fully explored the double-fading benefits of product channels. To address such a problem, in \cite{Liujing2022}, three high-performance methods, maximum reflecting signal-to-noise ratio (Max-RSNR), maximum ratio reflecting (MRR) and  selective ratio combining (SRR), were proposed. Their main advantages are low-complexity with closed-form or semi-closed-form solutions and realize some rate improvement over EGR in \cite{Dailinglong2021active}. \textcolor{blue}{ In active IRS aided single-input multiple-output scenario \cite{long2021active}, successive convex approximation (SCA) was proposed to separately design the amplitude and phase of the active IRS via maximizing user's achievable rate.} But, their aim is to not maximize  the total SNR at user in as \cite{Liujing2022}. For example, the best beamformer Max-RSNR in \cite{Liujing2022} is only to maximize the reflected SNR by omitting the direct-path signal. Thus, it  is  obvious that there is still a large performance room to improve the achievable rate in such an active IRS-aided system. Thus, in this paper, to harvest more rate gain achieved by  active IRS, we propose three enhanced-rate iterative beamformers with slight increased complexities.

\begin{enumerate}
\item To improve the achievable rate of methods in \cite{Dailinglong2021active}\cite{Liujing2022}, maximizing simplified SNR (Max-SSNR) is formalized.  Through omitting the cross-term, which will significantly simplify the solving process. Max-SSNR may be directly solved  by the distinct way:  Rayleigh-Ritz (RR) ratio, which are short for Max-SSNR-RR. Here, an iterative process is introduced between  the norm of beamforming vector and its normalized vector. In additon, motivated by the concepts of maximum ratio combining and Max-RSNR, an iterative generalized maximum ratio reflection (GMRR) is proposed. \textcolor{blue}{Simulation results show  that the two methods strike a good balance between rate performance and complexity.}
  \item To further improve the achievable rate, under the power constraint of active IRS, the objective function is to maximize the SNR. Two-level iteration by fractional programming (FP), which is called Max-SNR-FP. Simulation results show that the Max-SNR-FP method achieves about 1-3 bits higher than methods in \cite{Liujing2022} in terms of achievable rate, \textcolor{blue}{and is slightly better than alternative optimization (AO) in \cite{Dailinglong2021active} and  SCA in  \cite{long2021active} in terms of rate performance.}
      %They are in increasing order of rate performance: GMRR, Max-SSNR-RR and Max-SSNR-FP. It is particularly noted that the  Max-SSNR-FP achieves about 4-bit and 7-bit rate gains over Max-RSNR and passive IRS, respectively.
  %\item To further improve the achievable rate, with maximizing the reflected signal-to-noise ratio (Max-RNSR) as the objective function, the closed-form of the the beamformer is designed under satisfying the power constraint of IRS. An iterative generalized maximum ratio reflecting (GMRR) is proposed. The iterative optimized active IRS vector is obtained by iterating the normalized vector of the active IRS and the corresponding closed-form. Simulation results show that the Max-RSNR algorithm achieves 1-2 bits higher than EGR, MRR, and SRR in terms of achievable rate.
\end{enumerate}

The remainder is organized as follows. Section II presents the system model and three methods are proposed in Section III. In Section IV, numerical simulations are presented, and Section V draws conclusions.

$Notations$: In this paper, bold lowercase and uppercase letters represent vectors and matrices, respectively.
{Signs  $(\cdot)^H$, $(\cdot)^{-1}$, $\Re$, $|\cdot|$, $(\cdot)^*$ and $\|\cdot\|_2$ denote the conjugate transpose operation, inverse operation, real part operation,  modulus operation, conjugate operation and 2-norm operation, respectively. The notation $\textbf{I}_N$ is the $N\times N$ identity matrix, the sign $\mathbb{E}\{\cdot\}$ represents the expectation operation, and $\operatorname{Diag}(\mathbf{x})$ denotes the diagonal operator of converting vector $\mathbf{x}$ into a diagonal matrix.
\section{system model}
\begin{figure}[h]
\centering
\includegraphics[width=0.45\textwidth,height=0.23\textheight]{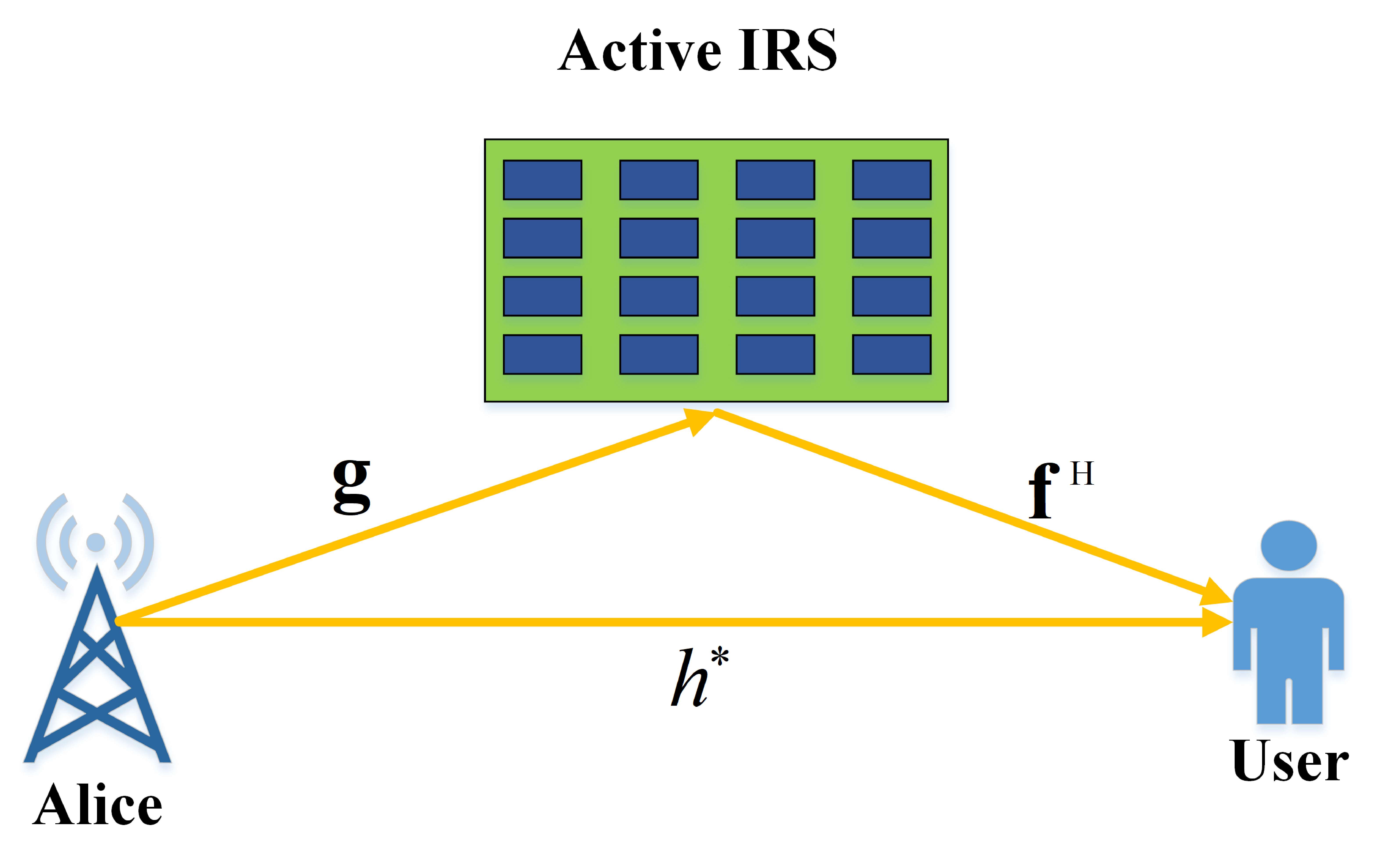}\\
\caption{System model diagram of an active IRS-aided wireless network.}\label{systemmodel.eps}
\end{figure}
Fig.~\ref{systemmodel.eps} depicts a diagram of an active IRS-assisted wireless network. Both base station (BS) and user are equipped with single antenna. IRS is equipped with $N$ active reflecting elements, and the amplification factor for  element $n$ is denoted by $p(n)$.
Without loss of generality, it is assumed that all three channels from the BS to the IRS, IRS to the user, and the BS to the user are Rayleigh fading.

The signal arriving at the IRS is given by
\begin{equation}
\mathbf{x}_{I}=\mathbf{g}s,
\end{equation}
where $\mathbf{g}\in\mathbb{C}^{N \times 1}$ stands for the channel from BS to IRS, and $s\in\mathbb{C}^{1 \times 1}$ indicates that the transmit signal at BS satisfying $E\left[|s|^2\right]=P_S$.

The signal reflected by IRS can written as
\begin{align}
\mathbf{y}_{I}=&\mathbf{P}\mathbf{x}_{I}+\mathbf{P}\mathbf{w}_{I}=\mathbf{P}\mathbf{g}s+\mathbf{P}\mathbf{w}_{I},
\end{align}
where  $\mathbf{P}=\operatorname{Diag}(\mathbf{p})$ denotes the diagonal matrix of the active IRS, and $\mathbf{w}_{I}\in\mathbb{C}^{N \times 1}$ represents the additive white Gaussian noise (AWGN) with distribution $\mathbf{w}_I \sim \mathcal{C N}\left(0, \sigma_I^2\right)$.

The total power reflected by IRS is given by
\begin{equation}\label{yi}
\begin{aligned}
E\left\{\left(\mathbf{y}_I\right)^H \mathbf{y}_I\right\} &=E\left(s^H s\right) \mathbf{g}^H \mathbf{P}^H \mathbf{P} \mathbf{g}+E\left(\left(\mathbf{P} \mathbf{w}_I\right)^H \mathbf{P} \mathbf{w}_I\right) \\
&=P_S \mathbf{p}^H \mathbf{G}^{H} \mathbf{G}\mathbf{p}+\sigma_I^2\|\mathbf{p}\|_2^2=P_{I},
\end{aligned}
\end{equation}
where $P_I$ represents the  reflecting power constraint of the IRS, and $\mathbf{G}=\operatorname{Diag}(\mathbf{g})$. Let us define a normalized form $\mathbf{\overline{p}}_{k}$ of the beamforming vector at IRS $\mathbf{p}$ as
\begin{equation}\label{p1}
\overline{\mathbf{p}}_k=\frac{\mathbf{p}}{\lambda_k},
\end{equation}
where $\lambda_{k}$=$\|\mathbf{p}\|_2$ with $k$ being the $k$th iteration in our iteration structure between the norm of $\mathbf{p}$ and its normalized vector.

Therefore, (\ref{yi}) can be rewritten as
\begin{equation}\label{PI}
E\left\{\left(y_I\right)^H y_I\right\}=\lambda_k^2 P_S \mathbf{\overline{p}}_k^H \mathbf{G} \mathbf{G}^H \mathbf{\overline{p}}_k+\lambda_k^2 \sigma_I^2\|\mathbf{\overline{p}}_k\|_2^2=P_{I}.
\end{equation}
According to (\ref{PI}), $\lambda_k$ can be given by
\begin{align}\label{lambda}
\lambda_k=\sqrt{\frac{P_I}{P_S \mathbf{\overline{p}}_k^H \mathbf{G} \mathbf{G}^H \mathbf{\overline{p}}_k+\sigma_I^2\|\mathbf{\overline{p}}_k\|_2^2}}.
\end{align}
The signal received by the user can be expressed as
\begin{equation}\label{yu}
y_{u}=h^*s+\mathbf{f}^{H}\mathbf{P}\mathbf{g}s+\mathbf{f}^{H}\mathbf{P}\mathbf{w}_{I}+w_{U},
\end{equation}
where $h^*\in\mathbb{C}^{1 \times 1}$, and $\mathbf{f}^{H}\in\mathbb{C}^{1 \times N}$ denotes the channels from BS to user and IRS to user, and $w_{U}$ indicates the AWGN at the user receiver.  According to (\ref{yu}), the total received power at user can be described as
\begin{align}\label{rp}
E\left\{\left(y_u\right)^H y_u\right\}=&P_S(|h^*+\mathbf{f}^{H}\mathbf{P}\mathbf{g}|^2)+\sigma_I^2\mathbf{f}^{H}\mathbf{P}\mathbf{P}^H\mathbf{f}+\sigma_{U}^2.
\end{align}

From (\ref{rp}), the signal-to-noise ratio (SNR) of the user can be formulated
\begin{equation}\label{SNR}
\mathrm{SNR}=P_S \frac{|h^*+\mathbf{f}^{H}\mathbf{P}\mathbf{g}|^2}{\sigma_I^2 \mathbf{f}^H \mathbf{P} \mathbf{P}^H\mathbf{f}+\sigma_U^2}.
\end{equation}
The achievable rate of the user can be written as
\begin{equation}
\begin{aligned}
&R=\log _2\left(1+\mathrm{SNR}\right) \\
&=\log_2\left(1+P_S \frac{|h^*+\mathbf{f}^{H}\mathbf{P}\mathbf{g}|^2}{\sigma_I^2 \mathbf{f}^H \mathbf{P P}^H \mathbf{f}+\sigma_U^2}\right).
\end{aligned}
\end{equation}

\section{Proposed Three Methods}
In what follows, to enhance the rate performance, three high-rate iterative methods, called Max-SSNR-RR, GMRR and Max-SNR-FP, are proposed for an active IRS-aided wireless network. Compared with passive IRS, MRR, SRR, EGR and Max-RSNR, the proposed three methods perform much better in terms of rate performance.

%The computational complexity of GMRR is
%\begin{align}
%\mathcal{O}(L_{1}(2N^3+13N-4))
%\end{align}
% float-point operations (FLOPs), where $L_{1}$ denotes the iterative numbers of optimization variables $\lambda_k$ and $\mathbf{\overline{p}}_k$.
\subsection{Proposed Max-SSNR-RR}
The total average receive  SNR at user is written as
\begin{equation}\label{SNR-1}
\mathrm{SNR}=P_S \frac{\mathbf{f}^H\mathbf{G}\mathbf{p}\mathbf{p}^H\mathbf{G}\mathbf{f}+2\Re\{\mathbf{f}^H\mathbf{G}\mathbf{p}h\}+h^*h}{\sigma_I^2 \mathbf{p}^H \mathbf{F} \mathbf{F}^H \mathbf{p}+\sigma_U^2}.
\end{equation}
Due to the fact that the cross-term $2\Re\{\mathbf{f}^H\mathbf{G}\mathbf{p}h\}$  in numerator in (\ref{SNR-1}) will lead to a difficulty of  optimizing $\mathbf{p}$ by maximizing SNR.  For convenience of calculating $\mathbf{p}$ below,  we set the cross term to $0$ to form a simplified signal-to-noise ratio (SSNR) as
\begin{equation}
\mathrm{SSNR}(\overline{\mathbf{p}}_k,~\lambda_k)=\frac{\overline{\mathbf{p}}_k^H\left(P_S \lambda_k^2 \mathbf{G}^H \mathbf{f f}^H \mathbf{G}+P_S h^* h \mathbf{I}_N\right) \overline{\mathbf{p}}_k}{\overline{\mathbf{p}}_k^H\left(\lambda_k^2 \sigma_I^2 \mathbf{F F}^H+\sigma_u^2 \mathbf{I}_N\right) \overline{\mathbf{p}}_k}.
\end{equation}
with replacing $\mathbf{p}$ in (\ref{SNR-1})  by $\overline{\mathbf{p}}_k\lambda_k$.
The optimization problem of maximizing SNR can be expressed as
\begin{subequations}
\begin{align}
&\max_{\mathbf{\overline{p}_k}}  ~~\mathrm{SSNR}(\overline{\mathbf{p}}_k,~\text{fixing}~\lambda_k)\\
~~&\text{s.t.}  ~~~\overline{\mathbf{p}}_k^H \overline{\mathbf{p}}_k=1.
\end{align}
\end{subequations}
 Accordingly, given the value of $\lambda_k$ and using the Rayleigh-Ritz ratio theorem, $\mathbf{\overline{p}}_k$ is the eigenvector corresponding to the largest eigenvalue of the matrix of the following
\begin{equation}\label{p}
\left(\lambda_k^2 \sigma_I^2 \mathbf{F} \mathbf{F}^H+\sigma_u^2 \mathbf{I}_N\right)^{-1} \left(P_S \lambda_k^2 \mathbf{G}^H \mathbf{f} \mathbf{f}^H \mathbf{G}+P_S h^*h \mathbf{I}_N\right).
\end{equation}
Plugging the solution $\mathbf{\overline{p}}_k$ to (\ref{p}) in (\ref{lambda}) gives  the value of $\lambda_k$ at iteration $k$. Then, an alternate iteration between $\lambda_k$ and $\mathbf{\overline{p}}_k$ are performed until the terminal condition $\lambda_k^{(p)}-\lambda_k^{(p-1)} \leq \epsilon$ is satisfied. The initial value of $\lambda_k$ with $k=0$ is assigned to be the MRR in \cite{Liujing2022}.

\subsection{Proposed GMMR}
From (\ref{SNR}), the reflecting SNR (RSNR) can be written as
\begin{equation}\label{RSNR}
\mathrm{RSNR}=P_S \frac{\mathbf{p}^H \mathbf{G}^H \mathbf{f f}^H \mathbf{G} \mathbf{p}}{\sigma_I^2 \mathbf{p}^H \mathbf{F} \mathbf{F}^H \mathbf{p}+\sigma_U^2}.
\end{equation}
Using the computed value of $\lambda_k$, the above RSNR expression can be converted into
\begin{equation}\label{RSNR1}
\mathrm{RSNR}=P_S \frac{\mathbf{p}^H \mathbf{G}^H \mathbf{f f}^H \mathbf{G} \mathbf{p}}{\mathbf{p}^H (\sigma_I^2  \mathbf{F} \mathbf{F}^H+\lambda_k^{-2}\sigma_U^2 \mathbf{I}_N) \mathbf{p}}.
\end{equation}
Let us define
\begin{align}\label{C}
\mathbf{C}(\lambda_k)=\left(\sigma_I^2\mathbf{F}\mathbf{F}^H+\lambda_k^{-2}\sigma_U^2\mathbf{I}_N\right)=\mathbf{D}(\lambda_k)\mathbf{D}(\lambda_k)^H,
\end{align}
which is called the feature matrix of Max-RSNR.  According to (\ref{C}), (\ref{RSNR1}) can be rewritten as
\begin{equation}\label{RSNR22}
\mathrm{RSNR}=P_S \frac{\mathbf{p}^H \mathbf{G}^H \mathbf{f f}^H \mathbf{G}\mathbf{p}}{\mathbf{p}^H \mathbf{D}(\lambda_k)\mathbf{D}(\lambda_k)^H \mathbf{p}}.
\end{equation}
Let us define
\begin{align}\label{P-1}
\hat{\mathbf{p}}=\mathbf{D}(\lambda_k)^H\mathbf{p}.
\end{align}
Therefore, (\ref{RSNR22}) can be rewritten as
\begin{equation}\label{RSNR2}
\mathrm{RSNR}=P_S \frac{\hat{\mathbf{p}}^H \mathbf{D}(\lambda_k)^{-1}\mathbf{G}^H \mathbf{f f}^H \mathbf{G} (\mathbf{D}(\lambda_k)^{-1})^{H} \hat{\mathbf{p}}}{\hat{\mathbf{p}}^H\hat{\mathbf{p}}}.
\end{equation}
Clearly, $\mathbf{C}(\lambda_k)$ is a diagonal matrix. Therefore, the eigenvalue decomposition form of $\mathbf{C}(\lambda_k)$ can be easily written as $\mathbf{I}_{N}\mathbf{\Sigma}\mathbf{I}_{N}$. $\mathbf{\Sigma}$ is written as follows:
%The eigen-value decomposition of $\mathbf{C}(\lambda_k)$  is easy to to compute its inverse square root costs a huge computational amount. To reduce this computational load,  all off-diagonal elements of matrix $\mathbf{C}(\lambda_k)$ being equal to zeros gives a simple diagonal matrix form $\mathbf{E}$
%\begin{align} \label{F_rf}
%\mathbf{E}=
%\left[
%\begin{array}{ccc}
%    \sigma_I^2\|f(1)\|^2+\frac{\sigma_U^2}{\lambda_k^{2}} & \cdots & 0 \\
%     \vdots   &  \ddots & \vdots \\
%    0 & \cdots & \sigma_I^2\|f(N)\|^2+\frac{\sigma_U^2}{\lambda_k^{2}}
%\end{array}

%\begin{equation}
%\mathbf{E}=\operatorname{Diag}\left(\left[{\sigma_I^2\|f(1)\|^2+\frac{\sigma_U^2}{\lambda_k^2}},\cdots,{\sigma_I^2\|f(N)\|^2+\frac{\sigma_U^2}{\lambda_k^2}}\right]\right), n=1, \cdots , N,  ~~~~(17) \nonumber
%\end{equation}

%\right ],
%\end{align}
%\begin{align}\label{F_rf}
%\textcolor{blue}{\mathbf{a}=\left[{\sigma_I^2\|f(1)\|^2+\frac{\sigma_U^2}{\lambda_k^2}},\cdots,{\sigma_I^2\|f(N)\|^2+\frac{\sigma_U^2}{\lambda_k^2}}\right]},n=1, \cdots , N,
%\end{align}

\begin{align}\label{F_rf}
\mathbf{\Sigma}=\operatorname{Diag}\left(\left[{\sigma_I^2\|f(1)\|^2+\frac{\sigma_U^2}{\lambda_k^2}},\cdots,{\sigma_I^2\|f(N)\|^2+\frac{\sigma_U^2}{\lambda_k^2}}\right]\right),
\end{align}
$n=1, \cdots , N,$ which directly gives the corresponding value of diagonal matrix $\mathbf{D}(\lambda_k)$

\begin{equation}
\mathbf{D}(\lambda_k)=\operatorname{Diag}\left(\mathbf{a}\right), n=1, \cdots , N,
\end{equation}
where $\mathbf{a}=\left[{\sqrt{\sigma_I^2\|f(1)\|^2+\frac{\sigma_U^2}{\lambda_k^2}}},\cdots,\sqrt{{\sigma_I^2\|f(N)\|^2+\frac{\sigma_U^2}{\lambda_k^2}}}\right]$.

%\begin{align} \label{F_rf}
%\textbf{D}=
%\left[
%\begin{array}{ccc}
%    \sqrt{\sigma_I^2\|f(1)\|^2+\frac{\sigma_U^2}{\lambda_k^{2}}} & \cdots & 0 \\
%     \vdots   &  \ddots & \vdots \\
%    0 & \cdots & \sqrt{\sigma_I^2\|f(N)\|^2+\frac{\sigma_U^2}{\lambda_k^{2}}}
%\end{array}
%\right ].
%\end{align}
Inspired by maximum ratio combining at receiver, the beamforming form is given by
\begin{align}
\hat{\mathbf{p}}=\frac{\mathbf{D}(\lambda_k)^{-1}\mathbf{G}^H \mathbf{f}}{\|\mathbf{D}(\lambda_k)^{-1}\mathbf{G}^H \mathbf{f}\|_2}.
\end{align}
Plugging the above expression in (\ref{P-1}) yields
\begin{align}\label{pn}
\mathbf{p}&=(\mathbf{D}(\lambda_k)^{-1})^H\hat{\mathbf{p}}\nonumber\\
&=(\mathbf{D}(\lambda_k)^{-1})^H\frac{\mathbf{D}(\lambda_k)^{-1}\mathbf{G}^H \mathbf{f}}{\|\mathbf{D}(\lambda_k)^{-1}\mathbf{G}^H \mathbf{f}\|_2}\cdot e^ {-j\phi_h},
\end{align}
%\begin{align}\label{pp}
%\mathbf{p}(n)&=(\mathbf{D}(n))^{-1}\hat{\mathbf{p}}(n)\nonumber\\
%&=\frac{1}{\sqrt{\sigma_I^2\|f(n)\|^2+\frac{\sigma_u^2}{\lambda^{2}}}}\frac{\mathbf{g}^H(n)\mathbf{f}(n)}{\|\mathbf{g}^H(n)\mathbf{f}(n)\|_2}
%\end{align}
where $\phi_h$ represents the phase of the direction channel complex gain $h$ from BS to user.

According to (\ref{lambda}) and (\ref{pn}), the iterations of $\lambda_k$ and $\mathbf{\overline{p}}_k$ are performed to reach convergence.

%The complexity of Max-SSNR-RR is
%\begin{align}
%\mathcal{O}(L_{2}(2N^3+6N^2+7N-1))
%\end{align}
%FLOPs, where $L_{2}$ denotes the iterative numbers of optimization variables $\lambda_k$ and $\mathbf{\overline{p}}_k$.
\subsection{Proposed Max-SNR-FP}
The SNR is rewritten as the following equation
\begin{subequations}\label{cvx}
\begin{align}
&\max_{\mathbf{p}}  ~~\frac{\mathbf{p}^H  \mathbf{A}(\lambda_k) \mathbf{p}+2P_S\Re\{\mathbf{f}^H\mathbf{G}\mathbf{p}h\}}{\mathbf{p}^H \mathbf{B}(\lambda_k) \mathbf{p}}\\
~&\text{s.t.}  ~~~\mathbf{p}^H \mathbf{C}\mathbf{p}\leq P_I,
\end{align}
\end{subequations}
where
\begin{align}
&\mathbf{A}(\lambda_k)=\left(P_S  \mathbf{G}^H \mathbf{f} \mathbf{f}^H \mathbf{G}+\lambda_k^{-2} P_S h^* h \mathbf{I}_N\right), \\
&\mathbf{B}(\lambda_k)= \left(\sigma_I^2 \mathbf{F} \mathbf{F}^H+\lambda_k^{-2}\sigma_u^2 \mathbf{I}_N\right),  \mathbf{C}= \left(P_s \mathbf{G}^H \mathbf{G}+\sigma_I^2 \mathbf{I}_N\right).
\end{align}
Which is  converted into
\begin{subequations}\label{fpp}
\begin{align}
\underset{\mathbf{p}}{\operatorname{max}} ~~& z(\mathbf{p},\lambda_k)-y w(\mathbf{p},\lambda_k) \\
\text { s.t. } ~& \mathbf{p}^H \mathbf{C}\mathbf{p}\leq P_I,
\end{align}
\end{subequations}
in terms of the Dinkelbach's Transform in \cite{shen2018fractional}, where $z(\mathbf{p},\lambda_k)=\mathbf{p}^H \mathbf{A}(\lambda_k) \mathbf{p} +2P_S\Re\{\mathbf{f}^H\mathbf{G}\mathbf{p}h\}$, $w(\mathbf{p},\lambda_k)=\mathbf{p}^H \mathbf{B}(\lambda_k) \mathbf{p}$, and $y$ is a new auxiliary variable.  The above optimization problem is addressed by the following iterative updating process
\begin{equation}\label{FP-A}
y[t+1]=\frac{z(\mathbf{p}[t])}{w(\mathbf{p}[t])},
\end{equation}
where $t$ is the iteration index.

In (\ref{fpp}), $z(\mathbf{p},\lambda_k)$ is a  non-concave function, which can be linearized as the convex function
\begin{equation}\label{FP-B}
2\Re(\mathbf{p}^H\mathbf{A}(\lambda_k)\mathbf{p}_0)-(\mathbf{p}_0^H\mathbf{A}(\lambda_k)\mathbf{p}_0)+2P_S\Re\{\mathbf{f}^H\mathbf{G}\mathbf{p}h\}.
\end{equation}
Placing the above linear approximation in  (\ref{fpp}) forms a convex optimization
\begin{subequations}\label{ppp}
\begin{align}
\underset{\mathbf{p}}{\operatorname{max}} ~~& 2\Re(\mathbf{p}^H\mathbf{A}(\lambda_k)\mathbf{p}_0)-(\mathbf{p}_0^H\mathbf{A}(\lambda_k)\mathbf{p}_0)+\nonumber\\
&2P_S\Re\{\mathbf{f}^H\mathbf{G}\mathbf{p}h\}-y[t+1] w(\mathbf{p}) \\
\text { s.t. } ~& \mathbf{p}^H \mathbf{C}\mathbf{p}\leq P_I,
\end{align}
\end{subequations}
which may be solved and implemented by using CVX to output the value of $\mathbf{p}(t+1)$, where  $\mathbf{p}_0$ is taken to be $\mathbf{p}_0=\mathbf{p}_{MRR}$ in \cite{Liujing2022}. Based on (\ref{lambda}),  (\ref{FP-A}), and (\ref{FP-B}),  a two-layer iterative structure is proposed as listed in the following table Algorithm I, named Max-SNR-FP.
%\begin{table}[htbp]
% \centering
% %\caption{General Power Iterative Alorithm}
% \label{tab:pagenum}
% \begin{tabular}{llll}
%  \toprule
%   \textbf{Algorithm I} Proposed Max-SSNR-FP method\\
%   %Using the Max-SR Rule\\
%  \midrule
%  Step 1: Select $\lambda_{MRR}$ in \cite{Liujing2022} as the initial value of $\lambda_0$, \\$\mathbf{p}_0=\mathbf{p}_{MRR}$, $k$ = 0. \\
%  Step 2:  Set $k$=0, and the value of terminal threshold $\epsilon_1$, $\epsilon_2$ \\
%  Step 3:  \textbf{repeat}\\
%  Step 4:  $\textcolor{blue}{t}$ = 0, $\mathbf{p}_k=\mathbf{p}_{MRR}$, and compute $z(\mathbf{p},\lambda_k)$, $w(\mathbf{p},\lambda_k)$, \\
%  and the objective function of (\ref{ppp});\\
%  ~~Step 4-1:  \textbf{repeat}\\
%  %~~Step 5:    \\
%  ~~Step 4-2:   Compute $y(t+1)$ using $(\ref{FP-A})$;\\
%  ~~Step 4-3:   Calculate $\mathbf{p}(t+1)$ using $(\ref{FP-B})$; \\
%  ~~Step 4-4:   $\textcolor{blue}{t}$ = $\textcolor{blue}{t}$ + 1;\\
%  ~~Step 4-5: \textbf{until}  $\|\mathbf{p}(t+1)-\mathbf{p}(t)\| \leq \epsilon_1$.\\
%  Step 5: Optimize the optimal $\mathbf{p}(t+1)$ and normalized to obtain $\mathbf{\overline{p}}(t+1)$. \\
%  Step 6: Substitute $\mathbf{\overline{p}}(t+1)$ into (\ref{lambda}) to get $\lambda_{k}$.\\
%  Step 7: $k$ = $k$ + 1;\\
%  Step 8: \textbf{until}  $\lambda_k+1-\lambda_k \leq \epsilon_2$, and record the optimal value $\lambda_k+1$.\\
% % 10: $\bm{\Theta}^{(p)}$, $\mathbf{v}_{a}^{(p)}$, and $\mathbf{v}_{AN}^{(p)}$ are the optimal value, and $R_{s}^{(p)}$ \\
%%  is the optimal SR.\\
%  \bottomrule
% \end{tabular}
%\end{table}
\begin{algorithm}
\caption{Proposed Max-SNR-FP method}\label{algorithm gano}
\begin{algorithmic}[1]
\STATE Select $\lambda_{MRR}$ in \cite{Liujing2022} as the initial value of $\lambda_0$, $\mathbf{p}_0 = \mathbf{p}_{MRR}$, $k = 0$.
\STATE Set $k = 0$, and the value of terminal threshold $\epsilon_1$, $\epsilon_2$.
\REPEAT
\STATE $t = 0$, $\mathbf{p}_k = \mathbf{p}_{MRR}$, and compute $z(\mathbf{p},\lambda_k)$, $w(\mathbf{p},\lambda_k)$, and the objective function of (\ref{ppp});\\
\STATE   \textbf{repeat}\\
  %~~Step 5:    \\
\STATE     Compute $y(t+1)$ using $(\ref{FP-A})$, Calculate $\mathbf{p}(t+1)$ using $(\ref{FP-B})$, $t$ = $t + 1$;\\
\STATE   \textbf{until}  $\|\mathbf{p}(t+1)-\mathbf{p}(t)\| \leq \epsilon_1$.\\
\STATE Optimize the optimal $\mathbf{p}(t+1)$ and normalized to obtain $\mathbf{\overline{p}}(t+1)$.
\STATE Substitute $\mathbf{\overline{p}}(t+1)$ into (\ref{lambda}) to get $\lambda_{k}$.
\STATE $k = k + 1$;
\UNTIL $\lambda_{k+1}-\lambda_k \leq \epsilon_2$, and record the optimal value $\lambda_{k+1}$.
\end{algorithmic}
\end{algorithm}
\section{Simulation Results and Discussions}
In this section, numerical simulation results are conducted to evaluate the rate and convergent performance of our proposed methods. Simulation parameters are set as follows: $P_{s}$ = 20 dBm, $P_{I}$ = 35 dBm, and $\sigma_{I}^{2}=\sigma_{U}^{2}$= -70 dBm. The BS,  active IRS, and user are located at (0 m, 0 m), (80 m, 20 m), and (150 m, 0 m), respectively. The path loss at the distance $\bar{d}$ is modeled as $g(\bar{d})=\mathrm{PL}_0-10 \gamma \log _{10} \frac{\bar{d}}{d_0}$, where $\mathrm{PL}_0$ = -30dB represents the path loss reference distance $d_0$ = 1m, and $\gamma$ is the path loss exponent. The path loss exponents from BS to IRS, IRS to user, and BS to user are 2.3, 2.3, and 3.9 respectively.%$\epsilon_1$ = $\epsilon_2$ = $10^{-3}$ in Max-SSNR-FP method.
\begin{figure*}
 \setlength{\abovecaptionskip}{-5pt}
 \setlength{\belowcaptionskip}{-10pt}
 \centering
 \begin{minipage}[htbp]{0.33\linewidth}
  \centering
  \includegraphics[width=2.56in]{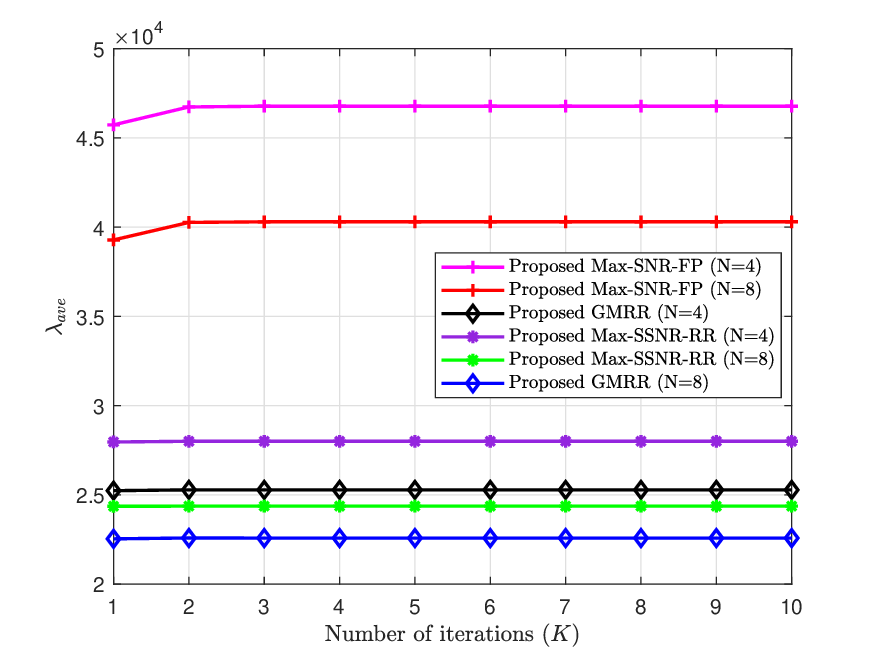}
  \caption{Convergent curves of proposed methods }
 \end{minipage}%
 \begin{minipage}[htbp]{0.33\linewidth}
  \centering
  \includegraphics[width=2.56in]{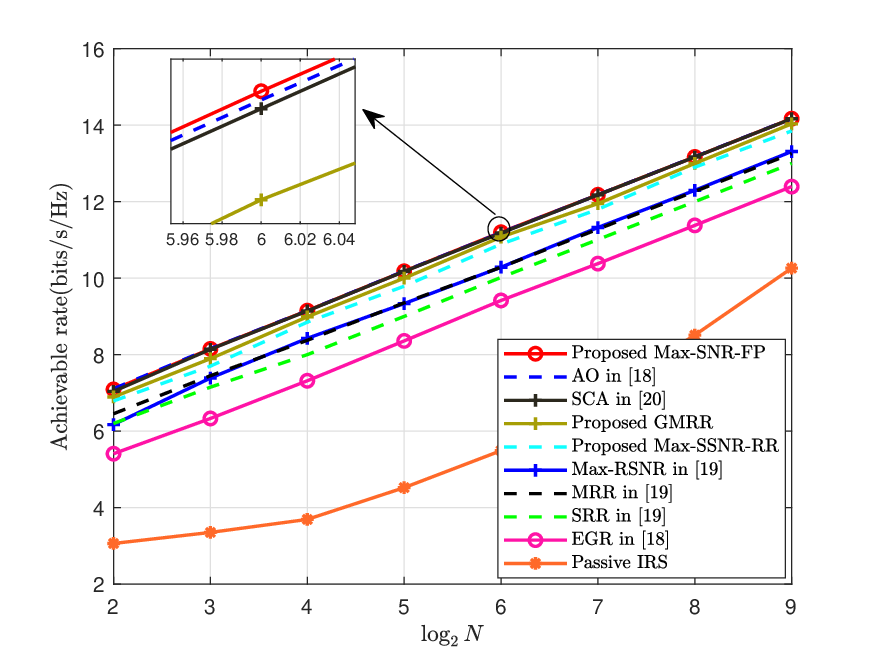}
  \caption{Achievable rate versus  $N$}
 \end{minipage}
 \begin{minipage}[htbp]{0.33\linewidth}
  \centering
  \includegraphics[width=2.56in]{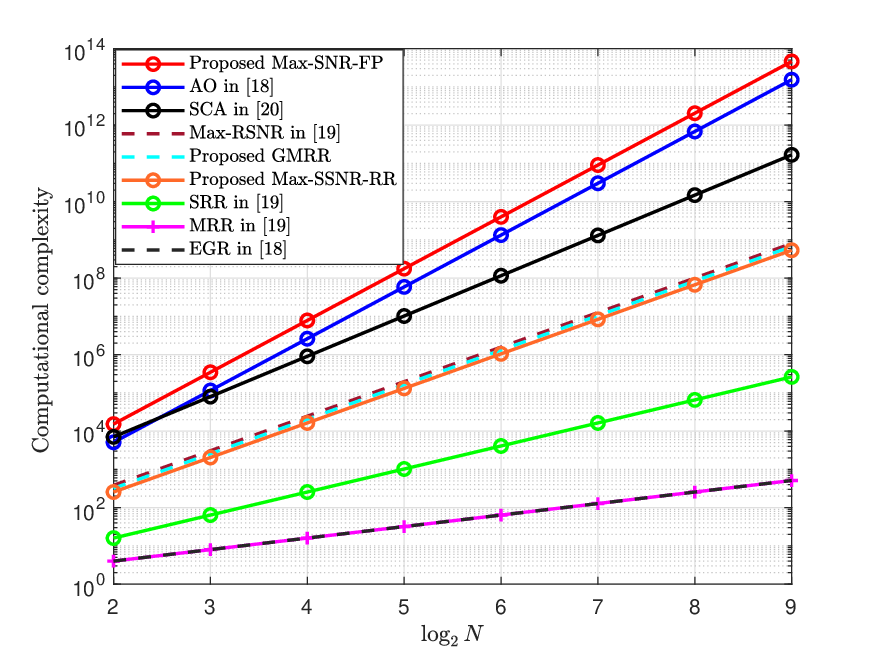}
  \caption{Computational complexity versus $N$}
 \end{minipage}
\end{figure*}

Fig.~2 depicts the convergent curves of the proposed three methods GMRR, Max-SNR-FP and Max-SSNR-RR with $N$ = 4 and 8. From the Fig.~2, it can be seen that all three methods proposed require about $3\sim5$ iterations to reach the convergence ceiling.
Therefore, they have a fast convergent speed.

%\begin{figure}[htbp]
%\centering
%\includegraphics[width=0.45\textwidth]{diedai.eps}\\
%\caption{Convergent curves of proposed three methods }\label{diedai.eps}
%\end{figure}
%\begin{figure}[htbp]
%\centering
%\includegraphics[width=0.45\textwidth]{RateN.eps}\\
%\caption{Achievable rate versus the number of IRS elements }\label{RateN.eps}
%\end{figure}
%\textcolor{blue}{Fig.~3 demonstrates the achievable rate versus the number of IRS elements for the proposed three methods Max-SSNR-RR, GMRR, and Max-SNR-FP  with all methods in \cite{Liujing2022},  algorithm 1 in \cite{Dailinglong2021active,long2021active} and passive IRS as performance comparison benchmarks.  The proposed three methods have made a significant rate enhancement over passive IRS. Particularly, the proposed Max-SNR-FP achieves about 4-bit rate gains over passive IRS when IRS is small-scale. Even, the proposed three methods still harvests more than 1-bit rate gains over the methods of MRR, SRR and EGR, and Max-SNR-FP  is comparable to that of algorithm 1 in \cite{Dailinglong2021active,long2021active}.}

\textcolor{blue}{Fig. 3 demonstrates the achievable rate versus the number of IRS elements for the proposed three methods Max-SSNR-RR, GMRR, and Max-SNR-FP with all methods in [19], AO in \cite{Dailinglong2021active}, SCA in \cite{long2021active} and passive IRS as performance comparison benchmarks. The proposed three methods have made a significant rate enhancement over passive IRS. Particularly, the proposed Max-SNR-FP achieves about 4-bit rate gains over passive IRS when IRS is small-scale. Even, the proposed three methods still harvest more than 1-bit rate gains over three closed-form solutions MRR, SRR and EGR. The proposed Max-SNR-FP is slightly better than AO in \cite{Dailinglong2021active} and SCA   in  \cite{long2021active} in terms of rate performance.}

\textcolor{blue}{Fig. 4 illustrates the computational complexities of the proposed methods versus the number of IRS elements with existing methods in \cite{Dailinglong2021active,Liujing2022,long2021active} as perfomance benchmarks. From this figure, it is seen that there is an increasing order in complexity (FLOPs): EGR ($\mathcal{O}(N)$ FLOPs), MRR ($\mathcal{O}(N)$ FLOPs), SRR ($\mathcal{O}(N^2)$ FLOPs), the proposed GMMR, the proposed Max-SSNR-RR, Max-RSNR in \cite{Liujing2022} ($\mathcal{O}(N^3)$ FLOPs), SCA in \cite{long2021active} ($\mathcal{O}(N^{3.5})$ FLOPs), AO in \cite{Dailinglong2021active} ($\mathcal{O}(N^{4.5})$ FLOPs), and Max-SNR-FP ($\mathcal{O}(N^{4.5})$ FLOPs). In paritcular, the proposed iterative GMMR and Max-SSNR-RR achieve far lower-complexity than AO and SCA in \cite{Dailinglong2021active,long2021active}, and AO in \cite{Dailinglong2021active} with a slight rate performance loss over the latter whereas the proposed Max-SNR-FP performs better than them using a higher complexity. Thus, we conclude the proposed GMMR and Max-SSNR-RR strike a good balance between rate performance and complexity.}

%\textcolor{blue}{Fig. 4 illustrates the computational complexities of the proposed methods versus the number of IRS elements with existing methods in \cite{Dailinglong2021active,Liujing2022,long2021active} as performance benchmarks. The proposed GMMR, Max-SSNR-RR, and Max-SNR-FP have computational complexities as follows: $\mathcal{O}(N^3)$, $\mathcal{O}(N^3)$, and $\mathcal{O}(N^4.5)$ float-point operations (FLOPs) while MMR, SRR, Max-RSNR in \cite{Liujing2022}, SCA in [20], and AO, EGR in \cite{Dailinglong2021active} are $\mathcal{O}(N)$, $\mathcal{O}(N^2)$, $\mathcal{O}(N^3)$, $\mathcal{O}(N^{3.5})$, and $\mathcal{O}(N^{4.5})$, $\mathcal{O}(N)$ FLOPs, respectively. In summary, the proposed iterative GMMR and Max-SSNR-RR achieve far lower-complexity than two existing methods AO and SCA in  \cite{Dailinglong2021active,long2021active}, and AO in \cite{Dailinglong2021active} with a slight rate performance loss over the latter whereas the proposed Max-SNR-FP performs better than existing methods using a higher complexity. Thus, the proposed GMMR and Max-SSNR-RR strike a good balance between rate performance and complexity.}

\section{Conclusions}
In this paper,  three high-performance methods, Max-SSNR-RR, GMRR, and Max-SNR-FP, were presented to separately optimize the amplitude and phase of active IRS  to make a rate enhancement for wireless network. They all adopt the same iterative architecture, which transforms the total beamforming problem at IRS into two subproblems: $(a)$ given its norm,  its phase or normalized beamforming vector was optimized by different rules like Max-SNR or Max-RSNR, $(b)$ given its normalized vector, its norm was computed by power constraint.   Simulation results indicated that the proposed three methods outperform MRR, SRR, EGR and the passive IRS case in accordance with rate. Their rates  are  in increasing order: Max-SSNR-RR, GMRR, and Max-SNR-FP. \textcolor{blue}{Additionally, Max-SNR-FP achieves a slightly higher achievable rate performance than  AO in \cite{Dailinglong2021active} and SCA  in \cite{long2021active}, and the proposed GMMR and Max-SSNR-RR have stricken a good balance between rate performance and complexity.}
\ifCLASSOPTIONcaptionsoff
\newpage
\fi

\bibliographystyle{IEEEtran}
\bibliography{IEEEfull,reference}

\end{document}